\documentclass[letterpaper,twocolumn,10pt]{article}

\usepackage{usenix}

\usepackage[T1]{fontenc}
\usepackage[utf8]{inputenc}
\usepackage{microtype}

\usepackage{amsmath,amssymb,amsthm,mathtools}

\usepackage{url}
\usepackage{appendix}
\usepackage{graphicx}
\usepackage{enumitem}
\graphicspath{{}}
\usepackage{pifont}
\usepackage{booktabs}
\usepackage{multirow}
\usepackage{array}
\usepackage{caption}
\usepackage{subcaption}

\usepackage{xcolor}

\usepackage[numbers,sort&compress]{natbib}

\usepackage[colorlinks=true,linkcolor=black,citecolor=black,urlcolor=blue,
            pdftitle={LiZoneDedup: FAST '27 submission}]{hyperref}
\usepackage[capitalize,noabbrev]{cleveref}

\usepackage{xspace}
\newcommand{\sysname}{LiZoneDedup\xspace}

\newcommand{\paragraphX}[1]{\vspace{2pt}\noindent\textbf{#1.}}
\providecommand{\email}[1]{{\itshape #1}}

\title{\sysname: An End-to-End In-Storage Deduplication\\
       Architecture for ZNS SSDs}

\author{
{\rm Fan Hao}\\
Information Support Force Engineering University\\
\email{haofan@isfeu.edu.cn}
\and
{\rm Xianwei Lv}\\
Information Support Force Engineering University\\
\email{xwlv@isfeu.edu.cn}
\and
{\rm Ruikun Luo}\\
HUST\\
\email{rkluo0924@gmail.com}
}

\date{}

\begin{document}
\maketitle


\begin{abstract}

Data deduplication reduces write traffic and extends flash lifetime, especially on versioned workloads: container images, VM snapshots, and backups. Yet in-storage deduplication has never been implemented on emerging Zoned Namespace (ZNS) SSDs. Conventional designs require a page-level FTL to remap duplicate writes and a device-wide fingerprint index far exceeding controller DRAM; neither exists under ZNS. Our insight is that ZNS's append-only contract is a natural substrate for deduplication: duplicate mappings, once recorded, are never modified in place, so dedup metadata becomes an append-only log rather than a mutable page-level table, and fingerprints fit in compact per-zone indexes rather than a global index. We present \sysname, the first complete in-storage deduplication architecture for ZNS, built around three mechanisms: a device-selected write primitive that hides zone
placement from the host, content-based routing that confines each lookup to a small fixed set of candidate zones, and lifecycle-separated metadata with per-zone identity indexes and append-only reference counts. On a 1\,TB ZNS SSD, \sysname replaces the $\sim$10.7 GB global fingerprint index with per-zone indexes of a few hundred KB that become NAND-resident once sealed, holds controller DRAM at 37--50 MB (including a bounded metadata LRU cache) regardless of device capacity, improves logical write throughput by 1.1--3.5$\times$, cuts physical writes by up to 85\%, and reaches 91--99\% of the throughput of an infeasible page-FTL oracle.

\end{abstract}

\section{Introduction}
\label{sec:intro}

Zoned Namespace (ZNS) SSDs are reshaping the flash storage stack by moving garbage collection, wear management, and fine-grained address translation out of the device~\cite{bjorling2021zns,tehrany2022understanding,bagashvili2025hidden}. On a 1\,TB device, the mapping table shrinks from hundreds of megabytes to a few hundred kilobytes, controller DRAM drops to 64--256\,MB, and write amplification becomes predictable---driving adoption across ZNS-aware file systems and storage stacks~\cite{kim2023raizn,li2024lsmraidzns,liu2024hizns}. Meanwhile, growing data volumes make it increasingly important to reduce the write traffic on ZNS SSDs: every NAND program suppressed translates directly into reduced wear, lower GC pressure, and longer endurance~\cite{dayan2015wamodel}.

In-storage data reduction has two complementary forms. Compression has been adapted to ZNS SSDs through slot-aligned designs~\cite{0Reviving,wang2024balloonzns}, yet its gain is content-dependent: compressed, encrypted, or random data yields little reduction, and every write pays computing costs. Deduplication eliminates entire redundant pages regardless of compressibility and is particularly effective on VM images, container layers, and backup streams~\cite{meyer2011study,jin2009effectiveness}. The two are stackable~\cite{saxena2022medes}, but deduplication remains absent from ZNS SSDs.

For conventional SSDs, deduplication runs either at the host or inside the device. Host-side dedup consumes gigabytes of host DRAM and CPU and is confined to one host or file system. In-storage dedup operates transparently before NAND programming, but every existing design was built for page-mapped SSDs and depends on two structures: a device-wide fingerprint index for duplicate detection and a page-level FTL for remapping a logical page to an existing physical copy~\cite{xu2025sampdedup, zhou2021remapssd, wu2022ead, srinivasan2012idedup}. Both require multi-gigabyte DRAM, far exceeding the 64--256\,MB controller budget of ZNS SSDs, so in-storage dedup cannot be directly applied to ZNS. Prior ZNS systems optimize zone allocation and reclamation, but none perform content-indexed duplicate detection; the only ZNS-aware dedup effort we know of, DeZNS~\cite{jamil2024dezns}, operates in the host file system and cannot suppress writes at the device. To our knowledge, \sysname is the first complete in-storage deduplication architecture designed for ZNS.


Our key insight is that ZNS removes the mutable page-level FTL that conventional dedup relies on, yet its append-only contract suits deduplication well: a duplicate relationship, once recorded, is never modified in place, so dedup metadata becomes an append-only log rather than a mutable table. \sysname introduces \texttt{dup\_append}, an extension of \texttt{zone\_append}: the host submits a payload without a destination zone, and the controller makes all placement and dedup decisions. Each request is resolved in three steps: (1)~\emph{content-derived routing} picks a small fixed set of candidate zones, each keeping its own small index, avoiding a global fingerprint table; (2)~\emph{zone-local lookup} verifies exact fingerprints within each candidate zone using split-state append-only metadata, with no in-place modification; (3)~\emph{reference-based hit recording} writes a 16-byte descriptor on a hit without page-level remapping, while a miss writes the full page. These mechanisms are backed by one safety guarantee: every cross-zone reference is counted, so a zone with outstanding references stays readable but closed to writes and is reset only after all references are released. \sysname is available at \url{https://anonymous.4open.science/r/lizonedup-50FD}.

We make three contributions:
\begin{itemize}[topsep=0pt]
\setlength{\itemsep}{0pt}
\setlength{\parsep}{0pt}
\setlength{\parskip}{0pt}
\item We identify the fundamental conflicts that make conventional in-storage deduplication infeasible on ZNS and derive the corresponding design requirements.
\item We design \sysname, an in-storage dedup architecture for ZNS combining
a device-selected \texttt{dup\_append} write primitive, content-derived zone
routing, and zone-local lookup with split-state append-native metadata.
\item We evaluate \sysname on five workloads. On a 1\,TB ZNS SSD, it reduces per-zone metadata to a few hundred KB, keeps controller DRAM at 37--50\,MB regardless of device capacity, improves write throughput by 1.1--3.5$\times$, and cuts physical writes by up to 85\% over standard ZNS.
\end{itemize}

This work focuses on the architectural feasibility of in-storage deduplication on ZNS SSDs: we resolve the fundamental conflicts that block conventional designs and validate correctness, metadata overhead, and performance under representative dedup-friendly, write-heavy workloads; corner-case stress tests and host-interface compatibility studies are deferred to future work (\S\ref{sec:conclusion}).

\section{Background and Challenges}
\label{sec:bg}

This section reviews ZNS properties and identifies three conflicts that make in-storage dedup infeasible.

\subsection{The ZNS Contract}
\label{sec:bg-zns}
\begin{figure}[b]

\centering
\subcaptionbox{Native ZNS SSD\label{znsori}}[0.24\textwidth]{%
  \includegraphics[width=0.10\textheight]{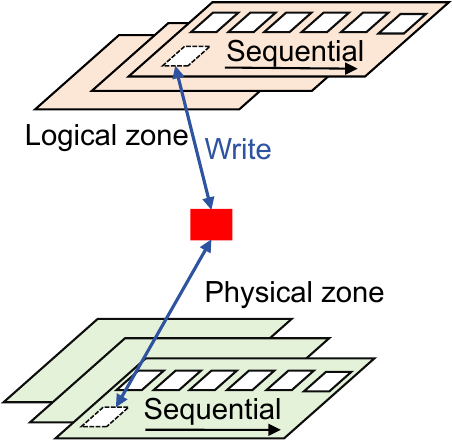}}%
\hfill
\subcaptionbox{Conventional dedup on ZNS\label{znsshe}}[0.24\textwidth]{%
  \includegraphics[width=0.14\textheight]{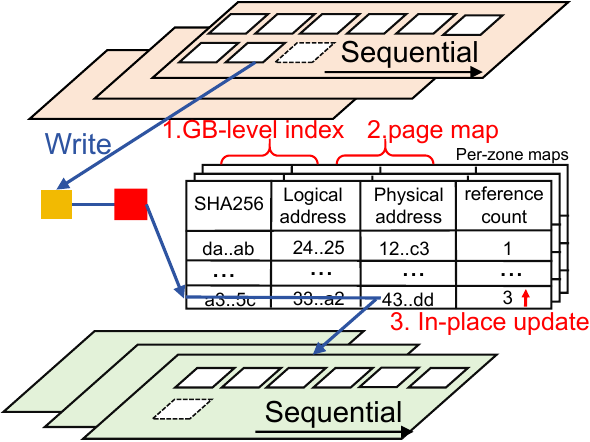}}
\captionsetup{justification=justified}

\caption{The mismatch between conventional in-storage deduplication and ZNS.
\emph{Red block} denotes the zone-level mapping table; \emph{yellow block}
denotes the hash generator.}
\label{dac}
\end{figure}

The NVMe Zoned Namespace (ZNS) command set partitions the logical address space into fixed-size zones (typically 64\,MB--2\,GB), each with a write pointer that advances strictly sequentially as~\Cref{znsori} shows. Unlike conventional SSDs, where the FTL absorbs out-of-place writes and lets the host address any \emph{Logical Block Address} (LBA) randomly, ZNS exposes this sequential constraint: writes must land at the current pointer, and an LBA cannot be overwritten until the whole zone is reset.

Reclamation requires a whole-zone \texttt{zone\_reset}, which erases the zone and resets its pointer. Only 16--32 zones may be open or active at once; a full zone transitions from appendable to sealed, and sealed zones stay readable but accept no writes until reset. The host manages placement and GC; the device is freed from page-level mapping and internal compaction. ZNS provides \texttt{zone\_append}~\cite{maheshwari2021blockstorocks} to simplify write pointer management: unlike \texttt{zone\_write}, which requires the host to specify both zone and LBA up front, \texttt{zone\_append} lets the host submit a zone and payload while the device picks the offset and returns the assigned LBA.

\subsection{Barriers to In-Storage Deduplication}
\label{sec:bg-fails}
Existing in-storage dedup designs assume a page-mapped FTL and a global fingerprint index~\cite{wu2022ead,zhou2021remapssd}, neither of which is available on ZNS. Three fundamental conflicts (\Cref{znsshe}) are detailed next.

\paragraphX{Conflict 1: A global fingerprint index vs.\ tiny controller DRAM} On a 1\,TB device with 4\,KB chunks, a page-level dedup index has roughly 268~million entries. At 40\,B per entry (32-byte SHA-256 fingerprint plus 8-byte address), the global index totals $\sim$10.7\,GB---two orders of magnitude larger than the 64--256\,MB DRAM on a contemporary ZNS controller~\cite{tehrany2022understanding, doekemeijer2023performance}. Partial caching and Bloom filters reduce the in-DRAM footprint, but not the core problem: the index is too large to fit in DRAM; even recent remapping-based approaches require 160--640 MB of dedicated NVRAM for a 256 GB device~\cite{zhou2021remapssd}, scaling to gigabytes at 1 TB.

\paragraphX{Conflict 2: Many-to-one duplicate mapping vs.\ zone-level-only translation} On a conventional SSD, a dedup hit updates one FTL entry in $O(1)$, remapping the host's logical page to the existing physical copy. ZNS keeps only a zone-level mapping ($\sim$8\,000 entries on 1\,TB), four orders of magnitude fewer. A dedup hit creates a many-to-one mapping. Multiple host-visible addresses resolve to the same physical page---precisely the page-level indirection ZNS removes. Without it, the standard ZNS interface cannot represent a dedup hit: with \texttt{zone\_write}, the host specifies the target LBA up front, so the device cannot decline to program after detecting a duplicate and return an already-occupied address; with \texttt{zone\_append}, the device selects the offset, but the command assumes every call writes new data and has no outcome where no full page is programmed yet a new LBA is returned.

\begin{figure}[t]
\centering
\includegraphics[width=0.92\linewidth]{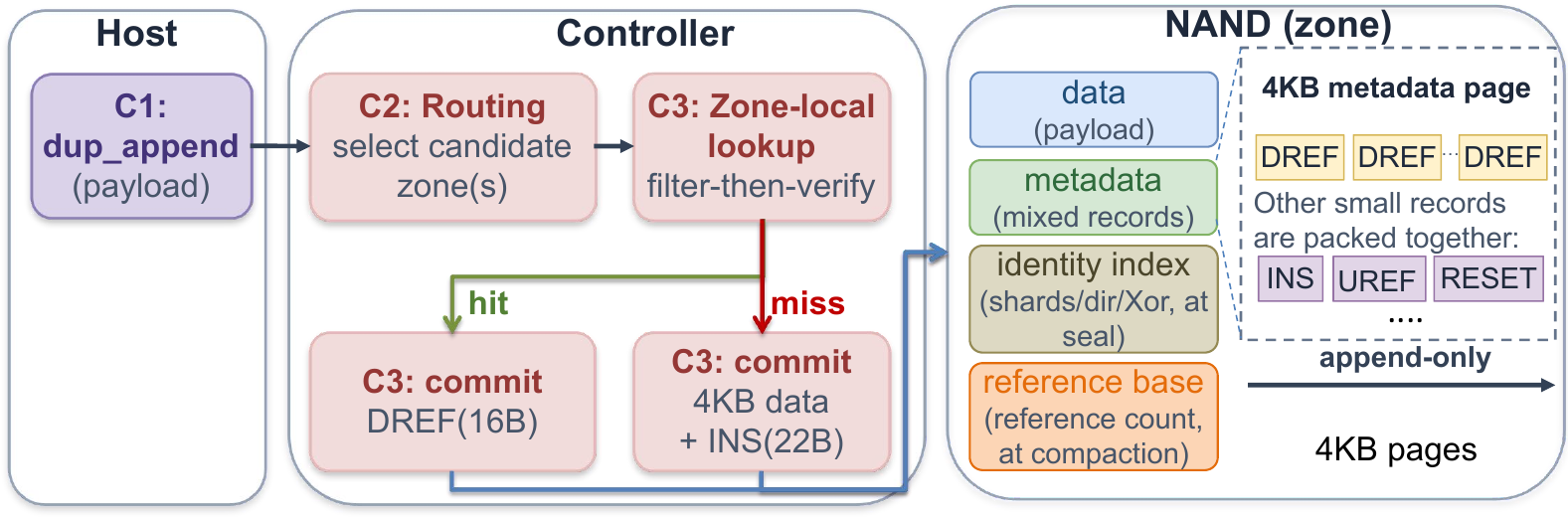}
\caption{Design Overview.}
\label{fig:znsoverview}
\end{figure}

\paragraphX{Conflict 3: Per-write metadata updates vs.\ append-only writes and tiny DRAM} Every write updates the fingerprint index. It inserts on a miss, reference-count increments on a hit, decrements on delete. On a conventional SSD, the FTL absorbs these updates transparently by writing a new physical page and updating the page-level map, so updates appear in-place to the host. On ZNS, this model breaks down. Writes are strictly append-only: without a page-level FTL to remap entries, existing metadata cannot be logically overwritten. Every update must be a new append, and random-writing a separate metadata region would break ZNS's predictable write amplification. At the same time, controller DRAM is tiny: any metadata that must stay resident across thousands of zones cannot fit in 64--256\,MB and would have to load on demand. Together, these constraints make conventional in-place-update metadata infeasible on ZNS.

\section{Design}
\label{sec:design}

This section presents \sysname in four parts: an overview, the \texttt{dup\_append} write primitive (C1), content-derived zone routing (C2), and zone-local lookup (C3).

\subsection{Design Overview}

The three conflicts of \S\ref{sec:bg-fails} stem from one fact: ZNS removes the page-level FTL that conventional dedup relies on. \sysname treats this constraint as a design premise. Its entry point is \texttt{dup\_append}, a device-selected append primitive modeled after \texttt{zone\_append}: the host submits a payload, and the controller makes all placement and dedup decisions. Deduplication operates on fixed 4\,KB page-aligned chunks. As shown in~\Cref{fig:znsoverview}, \sysname works as follows:

\begin{enumerate}[label=\textcircled{\arabic*}, itemsep=0pt, leftmargin=*,topsep=0pt]
\setlength{\itemsep}{0pt}
\setlength{\parsep}{0pt}
\setlength{\parskip}{0pt}
\item \textbf{Device-selected append (C1).} The host sends \texttt{dup\_append(payload)}; the controller selects both the destination zone and the offset, unlike \texttt{zone\_append} where the host names the zone.

\item \textbf{Content-derived routing (C2).} The device computes two values: a fingerprint for exact identity and a route digest for candidate-zone selection. The route digest picks a small constant set of $r$ candidate zones, serving as both lookup scope and allocation target pool, confining every foreground lookup to $O(r)$ zones.

\item \textbf{Zone-local lookup and commit (C3).} Within each candidate zone, the device probes a membership filter and, on a hit, reads one page-aligned shard for exact fingerprint verification. A match creates a 16-byte reference (possibly cross-zone) and returns the LBA without programming data; a miss appends the full data page to an appendable candidate and returns the new LBA. A failed lookup only means routing excluded the true duplicate zone (lowering dedup ratio, never correctness) or that the content is new.
\end{enumerate}

C3 realizes the metadata path for this flow: all mutations are sequential appends, immutable content identity is separated from mutable reference state, and per-zone indexes stay at hundreds-of-kilobyte scale, with DRAM growing with active-zone count rather than total device capacity.

\subsection{C1: The \texttt{dup\_append} Write Primitive}
\label{sec:design-c1}
Conventional in-storage dedup relies on two mechanisms that ZNS lacks. First, every dedup outcome requires a page-level mapping update---a miss allocates a new page, a hit remaps to an existing copy---but ZNS has only a zone-level map and no page-level FTL. Second, these updates are random and per-LBA, while ZNS writes are strictly append-only. \texttt{dup\_append} resolves both by moving placement and duplicate detection from the host to the controller, so that every dedup outcome is expressed as an append rather than a mapping update.

C1 centers on four core mechanisms---the command's durability semantics, the hit/miss execution flow, the append-only record format, and the safe cross-zone reset protocol---followed by supporting mechanisms: standard-ZNS compatibility, concurrency, and crash durability.

\paragraphX{Interface and Durability Semantics}
\texttt{dup\_append} extends \texttt{zone\_append} with controller-selected zone placement while preserving ZNS's sequential write-pointer advancement and post-write LBA readability. It can be realized as a new NVMe opcode or vendor extension. Standard writes and \texttt{dup\_append} may coexist; the host retains zone management authority. Formally, \texttt{dup\_append(payload)} $\rightarrow$ \texttt{lba}: the host submits a page-aligned payload without a target zone, and the controller completes hashing, lookup, and durable commit. The \emph{linearization point} is the durable persistence of either a full data page (miss) or a DREF record (hit). A crash before this point leaves no trace; after it the LBA is readable. This atomic commit point eliminates half-completed states. Crash-recovery details are in \S\ref{sec:design-c3}. Host software adapted for \texttt{dup\_append} submits payloads without target-zone specification; unmodified ZNS hosts can use a fallback built on standard \texttt{zone\_append}. In the fallback, the host pre-computes the route digest locally and hints preferred candidate zones to the controller via vendor-specific metadata; since the host retains placement authority, the controller loses full routing freedom, so the fallback can suffer a reduced deduplication hit rate and higher lookup overhead. The native \texttt{dup\_append} mode delivers full deduplication capability but requires both NVMe vendor-specific extensions and host-stack modifications. All our evaluations use the native primitive, which exposes the full capability of \sysname; quantifying the fallback path is left for future work.

\paragraphX{Hit/Miss Execution Flow}
For each incoming page payload, the controller performs bounded candidate zone lookup (detailed in \S\ref{sec:design-c2}) before commit. On a miss, the controller appends the full 4\,KB data page to an appendable (i.e., active) zone, appends an \texttt{INS} (Identity Insertion) record that registers the new fingerprint-to-location mapping, and returns the allocated LBA. On a hit, the controller creates a descriptor record instead of programming a physical data page. A descriptor is a compact 16-byte append-only record that occupies one logical slot but zero data-area page programs. Because descriptors may reference a source page in any candidate zone (active or sealed), every detected hit is recordable with no eager copy-forward (a read-path optimization for hot pages is described in \S\ref{sec:design-c3}).

\paragraphX{Descriptor Record Format and Metadata Logging}
Four append-only record types cover all metadata mutations: \texttt{INS} for identity insertion on a miss, \texttt{DREF} for duplicate reference creation on a hit, \texttt{UREF} for descriptor invalidation on discard or reset (a $-1$ delta), and \texttt{RESET\_INTENT} for generation-scoped slot invalidation at reset start.

A \texttt{DREF} is a 16-byte record with four fields: a descriptor slot ID, a source \emph{OID}, i.e., the stable object identifier $\langle$\texttt{zone id}, \texttt{generation}, \texttt{in-zone offset}$\rangle$, a fingerprint suffix, and a record-type byte. Multiple DREFs pack into one 4\,KB metadata page, so the amortized NAND write per hit is tens of bytes. A DREF always points to a data-backed source page, never to another descriptor, keeping read resolution to a single level of indirection. Its reference-count delta is embedded in the record; liveness reconstruction is in \S\ref{sec:design-c3}. A host discard emits a \texttt{UREF} but does not immediately reclaim the source page, which remains valid while any live descriptor references it.

\paragraphX{Cross-Zone Reference Tracking and Safe Reset Protocol}
The controller maintains an \emph{incoming-reference count} per zone: a lightweight 4-byte counter of how many descriptors in other zones reference pages in this zone. This supports cross-zone references without a global remapping table. All updates are persisted via append-only DREF/UREF records; DRAM holds the running aggregate. The aggregate requires four bytes per zone, approximately 32\,KB device-wide.

Zone reset must not invalidate data visible through a reference in another zone. A zone enters \emph{reset-pending}: it stops accepting appends and follows a strict ordered protocol: \ding{172}~C2 stops creating descriptors that source this generation; \ding{173}~the controller durably invalidates all host-visible slots via a \texttt{RESET\_INTENT} record; \ding{174}~the zone scans its descriptors and appends \texttt{UREF}s to release all outgoing references; \ding{175}~the zone waits until its incoming count reaches zero; \ding{176}~physical erasure proceeds and the generation advances.

This ordering is deadlock-free for any number of mutually referencing zones: a zone releases all of its outgoing references at step~\ding{174} before waiting at step~\ding{175}, so no zone ever waits on a reference that a waiting zone still holds, and the wait-for graph is acyclic. The key invariant is that a descriptor's outgoing reference is released only after the descriptor itself is no longer host-visible. Reset-pending zones occupy active-zone slots, so the controller keeps a small reserve pool of pre-opened appendable zones to replenish the allocation pool, staying within the typical 16--32 active-zone limit. Advanced dynamic tuning under extreme cross-zone reference load is left for future work.

\paragraph{Relationship to Standard ZNS.}
Unlike conventional in-FTL deduplication, which allows arbitrary LBA-to-physical-page remapping and demands a full page-level FTL, \sysname preserves native ZNS sequential zone mapping for data-backed LBAs. Descriptor-backed LBAs are resolved via zone-local metadata instead. Three host-visible deviations from standard ZNS exist: reads of descriptor-backed slots incur one extra indirection (\S\ref{sec:design-c3}), and both write and reset semantics are modified as described below.

\emph{Write.} A dedup hit writes only a metadata descriptor without programming a physical data page; zone capacity and write-pointer advancement follow standard ZNS rules. \sysname reduces physical NAND writes rather than logical capacity: each hit still consumes one logical slot, preserving ZNS host-visible accounting. Logical resources are reclaimed exclusively at zone granularity via \texttt{zone\_reset}.

\emph{Reset.}
The primary semantic addition is reset-pending: zones with unresolved cross-zone references stay readable but reject new writes until references drain. The host retains full \texttt{zone\_reset} and zone-lifetime authority. Cross-zone references impose host-GC overhead requiring reference resolution before reset, and host-initiated resets obey the five-step reference-count protocol. Our long-run evaluation uses aggressive-expiry background migration for low-valid-page sealed zones; host-side GC-traffic quantification is deferred to future work.


\paragraphX{Concurrency}
Two requests may write identical content at the same time; without coordination, one could create a descriptor pointing to a page the other has not finished writing or has already reclaimed. The controller therefore pairs source-page pinning with an in-flight fingerprint set. A source page is pinned from the moment its fingerprint is selected until its descriptor is committed, so it cannot be reclaimed mid-write. The controller also keeps a small DRAM set of fingerprints whose insertions are currently in flight; a concurrent request matching one of these fingerprints either waits for the first insertion to finish and then references it, or writes its own full page. The set grows only with the number of outstanding writes and needs no per-fingerprint lock.

\subsection{C2: Content-Derived Zone Routing}
\label{sec:design-c2}

Conventional device-wide fingerprint indexes scale linearly with SSD capacity, exceeding controller DRAM budgets. Content-derived zone routing solves this by mapping each incoming chunk to a fixed set of $r$ candidate zones, so every lookup probes only $O(r)$ zones instead of the entire device.

The candidate set is selected by a route digest, a content-derived value that affects only \emph{where to search} and never \emph{what counts as a match}; exact identity is verified separately by a fingerprint. Multiple candidates serve two purposes: they capture more duplicates, since similar content routes to adjacent zones and sealed sources age out of the primary slot, and they balance writes across allocation targets. Within each zone (detailed in \S\ref{sec:design-c3}), a filter-then-verify pipeline narrows the lookup further, replacing a gigabyte-scale device-wide index with per-zone indexes of a few hundred kilobytes each.

\begin{table*}[t]
\centering
\scriptsize
\setlength{\tabcolsep}{4pt}
\renewcommand{\arraystretch}{1.05}
\caption{LiZoneDedup metadata inventory.}
\label{tab:metadata}
\begin{tabular}{@{}p{4.5cm}p{0.9cm}p{4.3cm}p{5.8cm}@{}}
\toprule
\textbf{Structure} & \textbf{Plane} & \textbf{Residency / size} & \textbf{Lifecycle} \\
\midrule
\multicolumn{4}{@{}p{15.5cm}@{}}{\emph{Log records --- packed into 4\,KB log pages}} \\
INS & identity & per-zone delta log; $\sim$22\,B each & miss; folded into shards at seal \\
DREF $\cdot$ UREF $\cdot$ RESET\_INTENT & reference
  & device-wide log; 16--20\,B each, fixed
  & hit/discard/reset; folded into reference base at compaction \\
\midrule
\multicolumn{4}{@{}p{15.5cm}@{}}{\emph{Identity plane (per-zone)}} \\
working index; counting Bloom & identity & DRAM; active-zone
budget & zone open; discarded at seal \\
identity delta log & identity & DRAM buffer (small) + flash
& appended on miss; merged at seal\\
sorted shards; prefix dir.; Xor filter; root ptr & identity
  & flash; $\sim$400\,KB/zone ($\sim$20\,B/entry)
  & written at seal; read-only until erase \\
\midrule
\multicolumn{4}{@{}p{15.5cm}@{}}{\emph{Reference plane (device-wide)}} \\
metadata log segments & reference & flash; 4\,KB pages
  & append-only; reclaimed after compaction \\
ref-count accumulator & reference & DRAM; a few MB (bounded)
  & rolling; flushed periodically; rebuilt on crash \\
per-zone reference base & reference & flash; 4\,B/obj, $\le$128\,KB/zone
  & new immutable version at compaction \\
incoming-ref counts & reference & DRAM; 4\,B/zone ($\sim$32\,KB total)
  & zone lifetime; drain to 0 before erase \\
\midrule
\multicolumn{4}{@{}p{15.5cm}@{}}{\emph{Auxiliary (not a plane)}} \\
slot bitmap; segment dir.; redirect table & --- & flash; per-zone; a few KB each & zone lifetime \\
route table & --- & DRAM; $<$200\,KB & device lifetime \\
\bottomrule
\end{tabular}
\end{table*}

\paragraphX{Dual-Hash and Candidate Zone Selection}
Zone routing requires a content-derived value that groups similar content, while duplicate detection requires an exact fingerprint. The controller therefore computes two content-derived values for every chunk (the specific hash functions are implementation choices, not interface constraints): a \emph{content fingerprint} $F$ (truncated SHA-256, 12--16 B) for exact identity verification, and a \emph{route digest} $L$ (24-bit SimHash) for candidate-zone selection. With a 12-byte fingerprint and at most $2^{15}$ unique pages per 128\,MB zone, the probability that a lookup falsely matches any page in a zone is below $2^{-81}$, and the probability that any two pages within a zone collide (birthday bound) is below $2^{-67}$.

The route digest must be similarity-preserving: a cryptographic hash scatters similar content uniformly (avalanche effect), so fingerprint-prefix routing co-routes only byte-identical chunks and scatters slightly modified versioned content across unrelated zones. SimHash keeps versioned content within a small Hamming neighborhood, so most duplicates arrive while their source zone is still appendable. Routing errors never compromise correctness: false negatives only reduce recall (the chunk is written normally), and false positives are rejected by exact fingerprint comparison.

The high 16 bits of $L$ index an in-DRAM route table of $2^{16}$ entries ($\sim$192\,KB: two-byte zone id + one flag byte per entry), checkpointed periodically. Each bucket maps to a \emph{primary} zone; the candidate set is the primary plus $r{-}1$ adjacent buckets selected by Hamming distance on the high 16-bit prefix, for a total of $r$ candidates. With $r{=}2$ (default), the device searches the primary and one adjacent bucket; the low 8 bits of $L$, computed at insertion time, break ties and are not stored in the route table. The route table evolves with zone lifecycle to reconcile two conflicting goals: sealed zones must remain queryable for duplicate discovery, while new allocations must go to appendable zones. A sealed zone stays a lookup candidate but is demoted from allocation targets; sealed zones beyond the fixed $r$-candidate budget are not probed on the foreground path, keeping lookup cost constant at $O(r)$---this only reduces recall, never correctness. Each fresh appendable zone serves as primary for a contiguous range of buckets (typically hundreds); when it seals, those buckets are reassigned to a new appendable zone, so at any time $\le$16 active zones serve all $2^{16}$ buckets.

\paragraphX{Lookup Cost Bound}
The end-to-end lookup proceeds in three layers: candidate-zone selection (C2), per-zone membership filtering, and exact fingerprint comparison (the latter two detailed in \S\ref{sec:design-c3}). Foreground writes probe at most $r$ zones, and each zone's internal lookup costs at most one filter probe plus one 4\,KB shard read. A lookup never loads a sealed zone's entire $\sim$400\,KB index: it fetches only the membership filter (a few KB, one flash page) and, on a filter hit, the single 4\,KB shard containing the fingerprint prefix. Sealed-zone filters and shards load on demand into a bounded LRU cache, so the worst-case lookup cost is $r$ filter probes plus $r$ single-page reads---strictly $O(r)$, independent of device capacity.

\begin{figure}[t]
\centering
\includegraphics[width=0.92\linewidth]{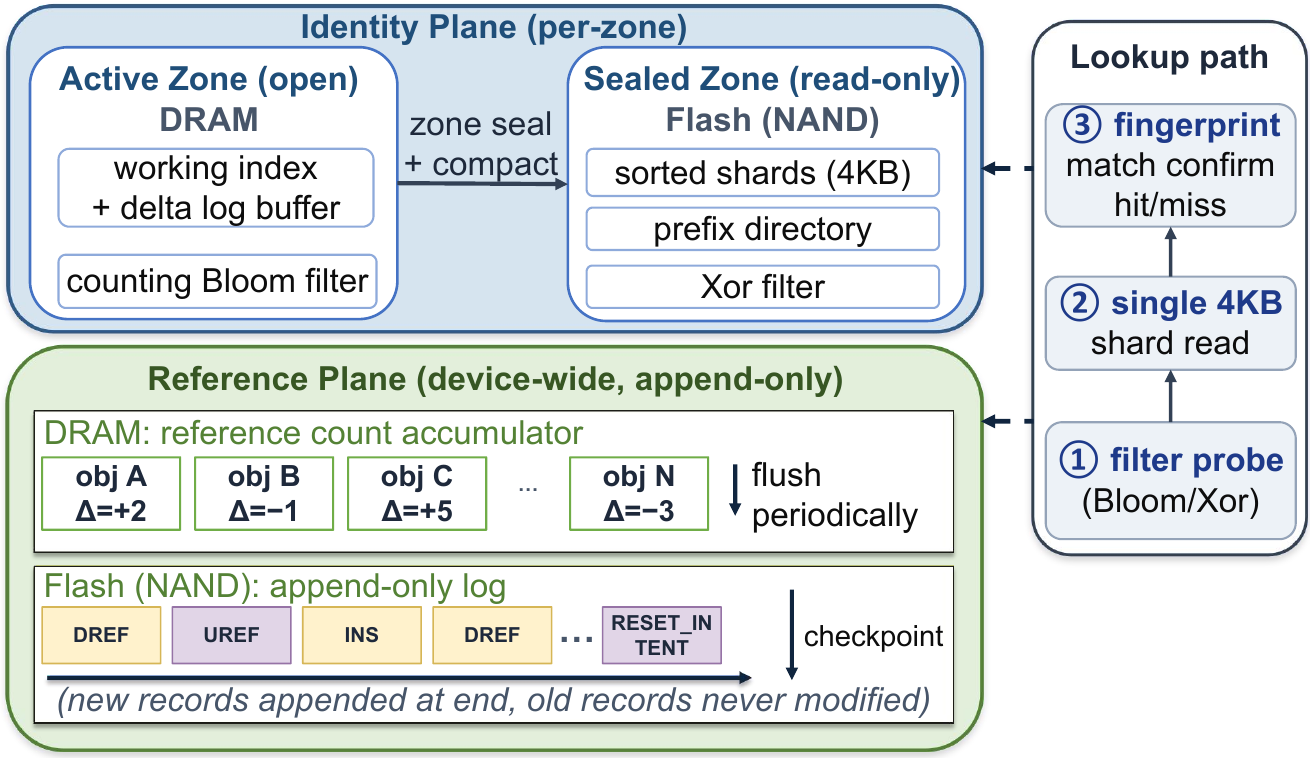}
\caption{Split-state metadata management.}

\label{fig:znsc3}
\end{figure}
\subsection{C3: Zone-Local Lookup and Split-State Metadata Management}
\label{sec:design-c3}

\sysname must satisfy three metadata constraints: small per-zone footprint, DRAM that scales with active-zone count rather than capacity, and append-only updates. As shown in~\Cref{fig:znsc3}, we meet all three with split-state separation along two axes. Along the lifecycle axis, content identity freezes at zone seal while reference liveness keeps evolving. Along the zone-state axis, active zones use insertion-friendly structures (counting Bloom filters, delta logs) and sealed zones use compact query-only structures (static Xor filters, sorted shards), keeping per-zone metadata small and DRAM limited to active zones.

\paragraphX{Split-State Architecture}
These two metadata types have fundamentally different lifecycles: a fingerprint-to-location mapping is written once on a miss and never changes, while reference counts evolve with every duplicate creation or discard. Mixing them prevents sealing a zone's lookup structures while references keep changing, and forces every reference update to rewrite identity data. We therefore separate them into two lifecycle-isolated planes: a lookup reads only the identity plane, and a hit appends only a reference record.

The \emph{identity plane} maps fingerprints to object identifiers (OIDs, \S\ref{sec:design-c1}); it is insertion-only while a zone is appendable and read-only after seal. The \emph{reference plane} records all reference-count updates, descriptor invalidations, and reset events exclusively as append-only log records (record types in \S\ref{sec:design-c1}); compaction writes new immutable bases and reclaims old segments without overwriting live data.

\paragraphX{Metadata Structures and Footprint}
\Cref{tab:metadata} lists every structure of both planes with its residency, size, and lifecycle. The identity plane comprises a DRAM working index and counting Bloom filter for active zones, an append-only identity delta log on flash, and, for sealed zones, 4 KB sorted shards with a prefix directory and a static Xor filter, all published through an identity root pointer at seal.  The reference plane comprises a device-wide append-only metadata log, a DRAM reference accumulator, a per-zone reference base, and per-zone incoming-reference counts. The per-zone identity index size is:
\begin{equation}
\text{IndexSize}_{\text{per-zone}} = N \cdot |\mathcal{E}|,
\label{eq:index-perzone}
\end{equation}
where $N$ is the number of unique pages per zone and $|\mathcal{E}|$ is
the per-entry size, independent of total device capacity. For a
representative 128\,MB zone with 4\,KB pages, assuming $N{=}20{,}000$ unique
pages, and $|\mathcal{E}|{=}20$\,B (a 12-byte truncated fingerprint and
an 8-byte zone-local offset, with the exact count kept in the reference
plane), \cref{eq:index-perzone} yields $\sim$400\,KB per zone. Sealed zones' indexes reside on NAND; only the $\le$16 active zones' working structures stay in DRAM, keeping controller DRAM at 37--50\,MB regardless of device capacity.

The steady-state DRAM working set totals $\sim$37--50\,MB: shard/filter caches dominate (tens of MB, configurable), active-zone metadata plus the reference accumulator add a few MB, and all other tables (route table, incoming counts, liveness summaries) total $<$200\,KB. Read-path auxiliary structures (per-slot status bitmap, segment directory, redirect table) are per-zone and bounded to a few kilobytes each (\emph{Read Path} below). Metadata caching follows three tiers: hot active zones keep all metadata in DRAM; warm active zones keep only the Bloom filter in DRAM and load shards on demand; sealed or cold zones reside entirely on flash with aggressive eviction after each query.

\paragraphX{Active and Sealed Identity State}
On a miss (new content), the controller writes the data page, assigns an OID, appends one identity record to the delta log, and updates the DRAM working index. Active-zone lookups probe the in-DRAM Bloom filter and working index (detailed in the Zone-Local Lookup Pipeline below), while sealed-zone lookups use the static Xor filter and sorted shards built at seal. The flash base shards are never rewritten on the foreground path. When a zone fills and seals, the controller stops accepting writes and performs a one-time seal-time compaction: it merges the delta log into sorted shards and builds a static Xor filter. The final step publishes a new identity root pointer. This publication is the crash-consistency boundary: a crash before it leaves the zone recoverable in its active (still-writable) form; a crash after it leaves a complete read-only sealed index. From this point on, all identity data is immutable: later duplicate hits only read it and send liveness changes to the reference plane.

\paragraphX{Zone-Local Lookup Pipeline}
Reading a full shard from NAND is expensive, so the lookup first uses a cheap membership filter to eliminate zones that definitely do not contain the fingerprint, then reads only the shards that may contain it. The end-to-end lookup is a three-layer pipeline. Layer~1 is candidate-zone selection (C2). Within each candidate zone, Layers~2 and~3 perform filter-then-verify: \emph{Layer 2---membership filter.} Each zone maintains an insertion-only counting Bloom filter while appendable and a static Xor filter once sealed. Both have false positives but no false negatives; a false positive costs one single-page shard read. Because deletion is zone-granular---an entire zone's filter and index are discarded together at reset---no per-key removal is ever required, which permits insertion-only structures. \emph{Layer 3---exact fingerprint comparison.} A prefix directory maps the fingerprint's high bits to a 4\,KB sorted shard, and a binary search within the shard confirms the match.

A foreground lookup reads at most one filter and one 4\,KB shard per candidate zone; for sealed zones, the Xor filter is fetched from the bounded cache or a single flash page, and the shard is similarly cached. Thus the per-zone lookup cost is one filter probe plus at most one single-page read, independent of the zone's total index size. To reduce cold-lookup latency, the controller employs two-stage prefetching. First, once the 24-bit SimHash route digest is available (before the full SHA-256 fingerprint completes), the controller asynchronously prefetches the Xor filter of each sealed candidate zone into DRAM; active-zone counting Bloom filters already reside in DRAM. Second, after the full fingerprint completes and a membership filter hits, the controller prefetches the target shard plus its two adjacent shards within the same zone, exploiting spatial locality. All prefetched data is managed by the bounded LRU cache.

\paragraphX{Read Path}
Reading a slot requires first determining whether it holds data or a descriptor, then locating the content accordingly. A slot is \emph{data-backed} when produced by a dedup miss (full 4\,KB page written), and \emph{descriptor-backed} when produced by a dedup hit (only a 16-byte DREF written, pointing to the existing source page).

For a data-backed slot, the read path is identical to standard ZNS: the controller checks the per-slot status bitmap, confirms the slot holds data, and the zone-level mapping directly locates the 4\,KB data page; one NAND read returns the data with no additional overhead. For a descriptor-backed slot, the controller checks the bitmap, confirms the slot holds a DREF, then computes the DREF's ordinal via \emph{rank-based local translation} (ranking the bitmap up to the target offset), uses a \emph{per-zone segment directory} to locate the DREF in the metadata log, extracts the source OID, and reads the source page, which may be in another zone. The cold path adds one metadata-log read before the source-page fetch; the hot path may serve both from DRAM cache.

For frequently-read descriptor-backed slots, the controller performs \emph{partial copy-forward}: the source page is copied to the local zone during idle periods, with a per-zone redirect entry mapping the old LBA to the new local page. Subsequent reads hit the in-DRAM redirect table and read locally, removing the indirection for hot slots. Cold slots are never copied, preserving dedup savings; the redirect table is per-zone, append-only, and bounded to a few hundred entries.

\paragraphX{Append-Only Reference Count Management}
The reference plane must record every reference-count change (duplicate hits, discards, resets) while keeping all writes append-only. It uses a device-wide append-only log (rather than per-zone logs) because descriptors reference pages across zones and records must be globally ordered for crash recovery. The log is organized into fixed-size segments, logically partitioned by owner zone; DREF and UREF records are appended on hits and discards, each carrying its source OID so the source zone is never modified even when sealed. When an owner zone resets, its records are invalidated by generation and segments are reclaimed lazily by segment cleaning. To reduce write amplification from hot objects, a bounded DRAM reference accumulator coalesces per-object count deltas before flushing; original DREFs remain in the durable log for recovery. Flushed deltas are periodically folded into per-zone reference bases via compaction (base + at most four unmerged segments, constant fan-in).

\paragraphX{Lazy Counting and Zone Reclamation}
\sysname must decide when to compute exact reference counts and when zones can be reclaimed, without scanning metadata on every update. Most reference updates do not need an immediate exact count: increments never make an object reclaimable, and many decrements leave multiple references. \sysname therefore delays exact count computation until it may affect reclamation; each zone keeps a coarse liveness summary for GC victim ranking, and exact state is computed only when an object may cross from one reference to zero or when its zone is selected for GC.

For GC-driven reclamation, a sealed zone with nonzero incoming-reference count (\S\ref{sec:design-c1}) is not selected for data-moving GC; only zones with zero incoming count are eligible for full reclamation, at which point no live descriptors reference their objects and no relocation is needed. This keeps the no-global-remapping invariant intact; the cost is that a zone may be pinned even when only a small fraction of pages remain referenced. To reclaim such pinned zones, the controller performs
\emph{background reference migration} on sealed zones with nonzero incoming count but valid-page fraction below 20\%: during idle periods, referenced pages are copied to a fresh zone. The controller then appends new DREFs pointing to the new location and UREFs invalidating the old ones. Referencing
zones need no modification---a read resolving an old DREF follows its UREF to the new copy. After migration the victim's incoming count reaches zero and it becomes eligible for reset. For host-initiated reset, the zone follows the five-step protocol of \S\ref{sec:design-c1}. During reset-pending, identity shards and the reference base remain readable for source-page resolution; only after final erasure are the zone's metadata structures discarded. Background reference migration enables reclamation of pinned sealed zones at the cost of extra physical NAND writes from copying live pages and rewriting reference descriptors. In the pathological worst case where nearly every zone becomes heavily cross-referenced, migration overhead can theoretically push total write amplification up toward 2$\times$, erasing a large fraction of deduplication-derived physical-write savings.

\paragraphX{Crash Recovery}
After a crash, the controller must reconstruct consistent metadata state. The linearization point defined in C1 (durable DREF or data-page commit) ensures no half-completed duplicate append survives a crash. On recovery, the controller (1)~replays the metadata log from the last checkpoint, reconstructing per-zone incoming counts, the reference accumulator, and active-zone working indexes from their delta logs; (2)~verifies each zone's identity root, falling back to the pre-seal active representation if seal did not complete; (3)~resumes any interrupted reset from its RESET\_INTENT record, applying each UREF at most once (each UREF names the unique DREF it releases, so recovery is idempotent); and (4)~rejects stale records carrying a previous zone generation. All recovery operations are append-only and require no in-place metadata fixes. The route table is checkpointed periodically to NAND: after a crash, the controller reconstructs it from the latest checkpoint and recovers post-checkpoint updates by scanning sealed-zone metadata, before serving host I/O.

\section{Evaluation}
\label{sec:eval}
We evaluate \sysname along five axes, ordered by importance: (1)~end-to-end bandwidth and latency (\S\ref{sec:eval-e2e}); (2)~data reduction effectiveness---dedup vs.\ compression vs.\ combination, and the impact of candidate-set size $r$ (\S\ref{sec:eval-reduction}); (3)~metadata read/write cost (\S\ref{sec:eval-metadata}); (4)~long-run 
behavior under cross-zone references (\S\ref{sec:eval-long}); and (5) scalability, zone-size sensitivity, and FEMU cross-validation~(\S\ref{sec:eval-sensitivity}).

\subsection{Methodology}
\label{sec:eval-method}

\paragraphX{Platforms}
All performance numbers come from an MQSim-style, event-driven, controller-level ZNS simulator~\cite{tavakkol2018mqsim}, modeling the full device pipeline: NAND page read 40\,\(\mu\)s, program 200\,\(\mu\)s, block erase 2\,ms (FEMU-default timings); 8 channels $\times$ 8 dies $\times$ 2 planes (128 plane-parallel units); a 4-core 2\,GHz controller; 128\,MB DRAM; the 16-zone active limit; and ZNS zone-state transitions.

\begin{table}[b]
\centering\small
\caption{Trace overview: redundancy and size. ``Oracle dedup'' is the reduction an exact global index achieves on the trace.}
\label{tab:traces}
\setlength{\tabcolsep}{3pt}
\scriptsize
\begin{tabular}{@{}lrrrrr@{}}
\toprule
\textbf{Trace} & \textbf{Writes} & \textbf{Logical} & \textbf{Unique} & \textbf{Dup.} & \textbf{Oracle} \\
 & (million) & (GB) & (GB) & frac. & dedup  \\
\midrule
FSL       & 3.38 & 12.9  & 10.5 & 0.19 & 1.23$\times$  \\
container  & 7.69 & 29.3  & 6.0  & 0.80 & 4.89$\times$  \\
VM         & 6.93 & 26.4  & 7.9  & 0.70 & 3.33$\times$  \\
backup     & 6.32 & 24.1  & 2.8  & 0.88 & 8.61$\times$  \\
rocksdb    & 1.85 & 7.1   & 5.9  & 0.17 & 1.20$\times$  \\
\bottomrule
\end{tabular}
\end{table}

Functional correctness is validated on a FEMU prototype~\cite{li2018femu} atop QEMU~10.1.0 (publicly available fork; provenance anonymized for review), with a 4\,GiB QLC device, 256\,MiB zones, and a ZenFS host stack; the prototype implements full zone-state transitions~\cite{shin2021znsplus,li2026warp}. Because FEMU runs under QEMU TCG without controller hardware parallelism, we compare only rank ordering and relative trends between simulator and FEMU~(\S\ref{sec:eval-sensitivity}). We inject crashes at five points around the linearization, seal, and reset protocols, four runs each (20 in total); every recovery yields zero dangling descriptors, idempotent UREF replay, and stale-generation rejection. Artifacts, including both simulator and the FEMU prototype, are publicly available at \url{https://anonymous.4open.science/r/lizonedup-50FD}. All simulation experiments are repeated at least three times; observed variance is small so error bars are omitted in plots.

\paragraphX{Workloads}
\Cref{tab:traces} summarizes the five write-dominated traces, which span the dedup-ratio spectrum and follow the standard workload methodology of prior deduplication and ZNS studies. Container~\cite{ng2011livededup}, VM~\cite{ng2011livededup,ng2013revdedup}, and backup traces~\cite{lillibridge2009sparse} are synthesized from real-world incremental snapshots to stress deduplication behavior; FSL is a raw real-world trace~\cite{ng2013revdedup,maheshwari2021blockstorocks}; rocksdb is the standard ZNS KV benchmark~\cite{maheshwari2021blockstorocks,bjorling2021zns}.

\paragraphX{Baselines}
Eight designs bracket the design space: \emph{NoDedup-ZNS} (standard ZNS); \emph{HostDedup-1GB} (host-side global fingerprint table in 1\,GB host DRAM, modeled as a software SHA-256 plus an in-DRAM hash lookup on the host before each write); \emph{PageFTLDedup}, the conventional in-storage design~\cite{xu2025sampdedup} ported to a hypothetical ZNS that retains a page-level FTL and an unconstrained in-DRAM fingerprint table, which is infeasible on real hardware (the $\sim$10.7\,GB index of \S\ref{sec:bg-fails}); \emph{\sysname} in software (LiZone-sw) and hardware (LiZone-hw) variants; \emph{Balloon-ZNS}~\cite{wang2024balloonzns}, in-storage LZ4 compression, likewise in software (Balloon-sw) and hardware (Balloon-hw) variants; and \emph{\sysname+Balloon}, stacking dedup and compression. DeZNS~\cite{jamil2024dezns} is not included because it deduplicates inside ZenFS with CRC-based prediction and cannot suppress device writes; HostDedup-1GB bounds what any host-side scheme can achieve.

\paragraphX{Configuration}
Unless otherwise noted: 1\,TB ZNS namespace, 128\,MB zones (8{,}192 zones), 4\,KB pages, 128\,MB controller DRAM, and $r{=}2$. We treat hardware acceleration as the default deployment configuration and software as the fallback, because SHA-256 acceleration is already standard on enterprise self-encrypting drives. A SHA engine computes the fingerprint in $\sim$2\,\(\mu\)s/page (8\,\(\mu\)s software)~\cite{intel2021ipp, zhou2021remapssd}; the SimHash route digest uses a non-cryptographic hash and runs on the controller core at 4\,\(\mu\)s, needing no hardware accelerator. The compression baseline uses LZ4 at 1.5\,\(\mu\)s/page (8/2\,\(\mu\)s compress/decompress in software)~\cite{wang2024balloonzns, lz4benchmark}.

\subsection{End-to-End Performance}
\label{sec:eval-e2e}

\begin{figure}[t]
\centering
\subcaptionbox{Write throughput.\label{fig:e2e-tput-write}}{%
  \includegraphics[height=0.12\textheight]{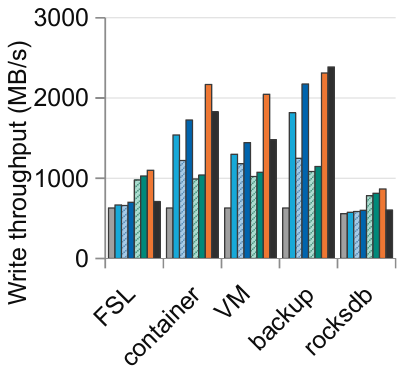}}%
\subcaptionbox{Read throughput.\label{fig:e2e-tput-read}}{%
  \includegraphics[height=0.12\textheight]{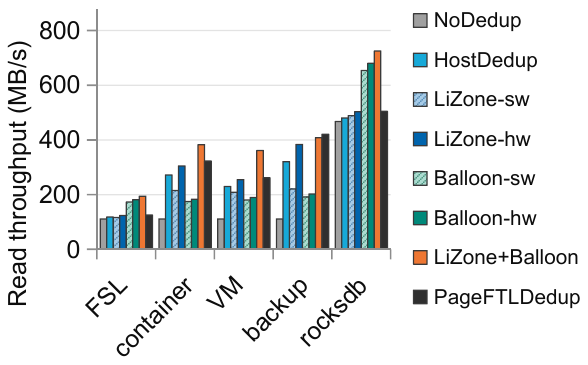}}
\caption{Throughput.}
\label{fig:e2e-thr}
\end{figure}

\paragraphX{Write throughput}
On dedup-rich workloads (\cref{fig:e2e-tput-write}), LiZone-hw raises logical write throughput over NoDedup by 2.7$\times$ on container, 2.3$\times$ on VM, and 3.5$\times$ on backup. Hits commit 16-byte descriptors instead of programming 4\,KB pages. It beats HostDedup-1GB by 12--20\% while consuming zero host CPU, and reaches 94.4\% of the infeasible PageFTLDedup oracle on container and 91.1\% on backup (97--99.8\% on dedup-light), since a miss follows the same hash-then-program path as the oracle, plus only microseconds of candidate-set lookup. Hardware acceleration matters because hashing is on every write's critical path: LiZone-sw is 29--43\% slower on dedup-rich workloads but at most 6\% slower on dedup-light.

Balloon-hw improves throughput by 1.65$\times$ on container and 1.8$\times$ on backup, but stays flat
(${\sim}$1.0--1.2\,GB/s) across workloads because LZ4 ratio tracks page compressibility rather than cross-version redundancy; it beats LiZone-hw on dedup-light workloads such as FSL. \sysname+Balloon achieves the highest write throughput on every workload, as dedup removes redundant pages while compression shrinks the remaining unique bytes.

On dedup-light workloads, LiZone-hw matches or slightly exceeds NoDedup: 1.11$\times$ on FSL and 1.08$\times$ on rocksdb, because the hash engine and occasional shard read cost less than the 200\,$\mu$s program they eliminate.

\begin{figure}[t]
\centering
\subcaptionbox{Write p99.\label{fig:e2e-lat-w99}}{%
  \includegraphics[height=0.128\textheight]{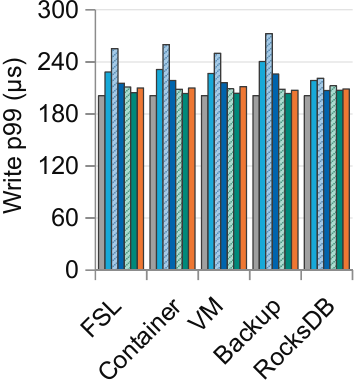}}%
\subcaptionbox{Write p50.\label{fig:e2e-lat-w50}}{%
  \includegraphics[height=0.128\textheight]{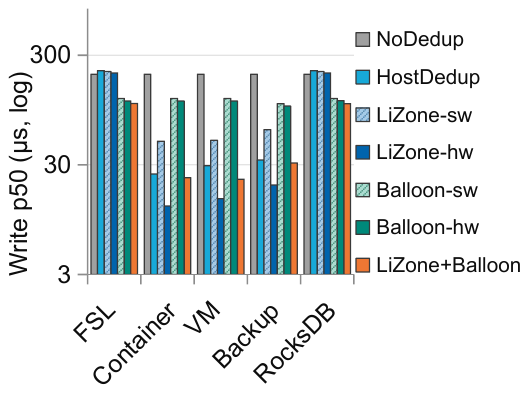}}
\caption{Write latency.}
\label{fig:e2e-lat-wr}
\end{figure}

\paragraphX{Write latency}
\Cref{fig:e2e-lat-w50,fig:e2e-lat-w99} show the write distribution. The P99 tail rises 4--13\% over NoDedup: the tail is a full-page program, identical to NoDedup's, plus the fingerprint and bounded lookup overhead. The median tracks dedup rate: on dedup-rich workloads the median write is a hit and P50 drops to 12.6--63\,$\mu$s (up to 16$\times$ below NoDedup's 201.0\,$\mu$s; 12.6--19.7\,$\mu$s in hardware, 49--63\,$\mu$s in software); on FSL and rocksdb the median is a miss and P50 matches NoDedup. The residual P99 overhead is fundamental: every cold miss still pays one 200\,$\mu$s program. Better miss-path pipelining is left to future work.

\begin{figure}[t]
\centering
\subcaptionbox{Read p99.\label{fig:e2e-lat-r99}}{%
  \includegraphics[height=0.128\textheight]{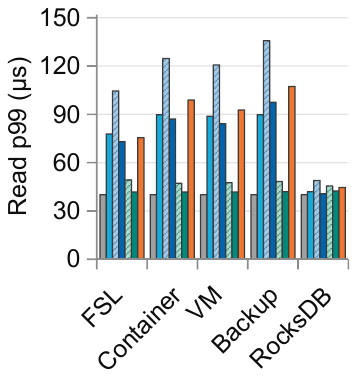}}
\subcaptionbox{Read p50.\label{fig:e2e-lat-r50}}{%
  \includegraphics[height=0.128\textheight]{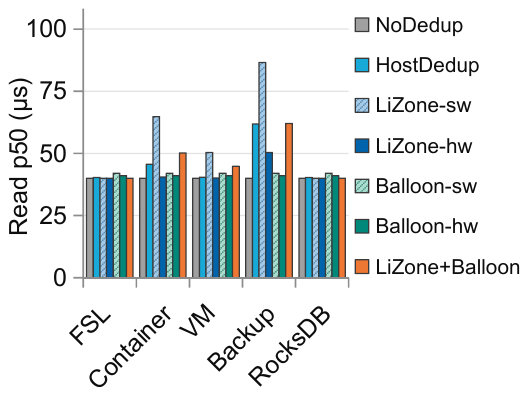}}%
\caption{Read latency.}
\label{fig:e2e-lat-re}
\end{figure}

\paragraphX{Read throughput}
On dedup-rich workloads (\cref{fig:e2e-tput-read}), LiZone-hw raises read throughput by reducing NAND interference: container reads rise from 111 to 305\,MB/s (2.7$\times$), VM to 255\,MB/s (2.3$\times$), and backup to 384\,MB/s (3.5$\times$), because fewer programs queued ahead of reads free more read slots on the plane schedulers. On dedup-light workloads, FSL reads stay at 124\,MB/s (112\% of NoDedup) and rocksdb at 503\,MB/s (108\%), since suppressed writes are too few to shift the interference profile. Balloon-ZNS removes less physical traffic than dedup on these workloads, so its interference reduction is smaller.

\paragraphX{Read latency}
Read P50 (\cref{fig:e2e-lat-r50}) stays at $\sim$40\,$\mu$s across all hardware variants: a data-page read is the same NAND operation regardless of dedup. The software variants (HostDedup, LiZone-sw) pay their hash and lookup path on the read side too, pushing P50 higher on dedup-rich workloads. The P99 tail (\cref{fig:e2e-lat-r99}) follows the same pattern: LiZone-hw adds a modest tail increase on dedup-rich workloads (container, VM, backup), the cost of one metadata-log page fetch for a cold descriptor before the source read (none when the DREF is cached). RocksDB is unchanged, as its reads hit the DRAM redirect path. Idle-time copy-forward mitigates this overhead for frequently accessed descriptors but makes little progress under sustained heavy I/O load: reduced physical NAND writes and flash longevity trade off against potential cold-read tail inflation.

\subsection{Data Reduction Effectiveness}
\label{sec:eval-reduction}

\begin{figure}[b]
\centering
\includegraphics[height=0.145\textheight]{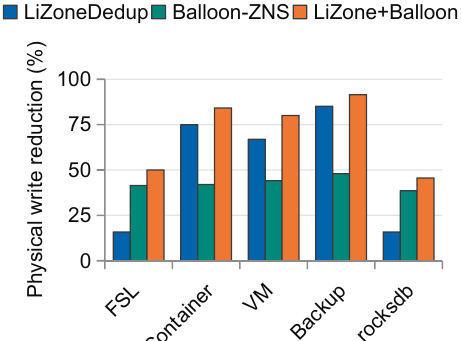}
\caption{Physical write reduction.}
\label{fig:dedup-reduction}
\end{figure}

\paragraphX{Dedup, compression, and their combination}
\Cref{fig:dedup-reduction} separates the two reduction mechanisms. \sysname eliminates 74.9\% of physical writes on container, 85.1\% on backup, and 66.9\% on VM, but only 15.8\% on FSL and 15.9\% on rocksdb. Balloon-ZNS removes 38.5--48.0\% with little workload sensitivity, since LZ4 finds some redundancy in nearly every page. The combination is additive on every workload, confirming the two layers stack without interference.

\paragraphX{Impact of $r$}
The candidate-set size $r$ trades data-reduction gain against metadata cost (\Cref{tab:reduction-r}). Reducing $r$ below 2 misses duplicates that route to adjacent buckets: reduction drops by 4--10 points, with the largest relative losses on the dedup-light FSL and rocksdb. Increasing $r$ beyond 2 yields only marginal gains (at most 4.5 points at $r{=}4$) while adding extra filter probes and shard reads per request. RocksDB saturates at $r{=}2$ because its duplicates are temporally adjacent compaction re-emissions. The default $r{=}2$ sits at the knee: it recovers most of the oracle's reduction with bounded lookup cost. Container and VM still leave 3--4.5 points on the table at $r{=}2$; adapting $r$ per workload from the observed hit rate is future work.

\begin{table}[t]
\centering\small
\caption{Physical write reduction (\%) under different $r$s.}
\label{tab:reduction-r}
\setlength{\tabcolsep}{4pt}
\scriptsize
\begin{tabular}{@{}lcccc@{}}
\toprule
\textbf{Workload} & \textbf{$r{=}1$} & \textbf{$r{=}2$} & \textbf{$r{=}3$} & \textbf{$r{=}4$}  \\
\midrule
FSL       & 6.9 & 15.8 & 18.2 & 18.8  \\
container & 69.3 & 74.9 & 78.5 & 79.4  \\
VM        & 59.7 & 66.9 & 69.8 & 70.0  \\
backup    & 80.9 & 85.1 & 87.1 & 87.1 \\
rocksdb   & 5.6  & 15.9 & 15.9 & 15.9  \\
\bottomrule
\end{tabular}
\end{table}

\begin{figure*}[t]
\centering
\subcaptionbox{Metadata read amplification.\label{fig:ablation1}}{\includegraphics[height=0.15\textheight]{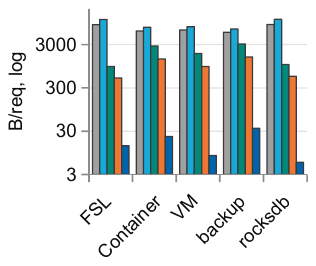}}%
\hfill
\subcaptionbox{Metadata write amplification.\label{fig:ablation2}}{%
  \includegraphics[height=0.15\textheight]{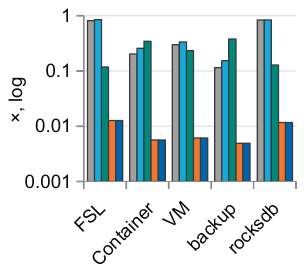}}%
\hfill
\subcaptionbox{Total write amplification across five layers; NoDedup = 1.0$\times$.\label{fig:ablation3}}{\includegraphics[height=0.15\textheight]{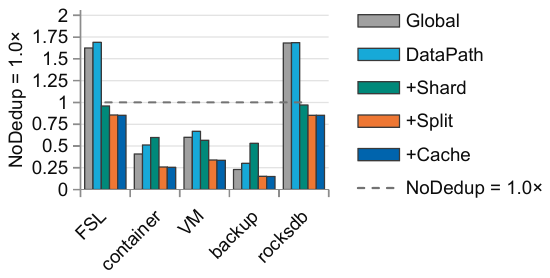}}
\caption{Amplification.}
\label{fig:metadata-ablation}
\end{figure*}

\subsection{Metadata Cost and Write Amplification}
\label{sec:eval-metadata}
This experiment answers whether \sysname eliminates the NAND I/O cost of exact-dedup indexes and preserves ZNS's low write amplification. We use a five-layer ablation: Global index (conventional device-wide hash table, the infeasible reference), Data path only (C1+C2 with a naive per-zone open-addressed hash table paged from NAND in 32-page chunks and updated in place), +Shard (page-aligned sorted shards with a prefix directory), +Split (the active/sealed lifecycle split with append-only delta logs), and +Full cache (complete \sysname with the three-tier DRAM cache).

\paragraphX{Metadata reads}
\Cref{fig:ablation1} shows metadata reads per request. The global-index layer costs 5.8--8.8\,KB, which is two orders of magnitude more than the ${\sim}20$\,B entry it consults. The open-addressed data path is no better (6.9--11.5\,KB): each lookup traverses open-addressed buckets spanning dozens of NAND pages. Page alignment (+Shard) cuts this to 0.9--3.1\,KB by making every post-filter confirmation a single 4\,KB read; the lifecycle split (+Split) contributes a further 1.6--2.0$\times$ via compact static filters on sealed zones. The DRAM cache collapses the residual to 5.5--34.9\,B, a 196--2{,}003$\times$ reduction versus the data path; on container, 99.7\% of confirmations hit the shard cache.

\paragraphX{Metadata writes}
Split-state separation makes metadata writes sequential (\Cref{fig:ablation2}). Relative to +Shard, which still rewrites shards in place, the append-only delta log cuts metadata write amplification by 61$\times$ on container and 11$\times$ on rocksdb; across all workloads the split-state plane runs at 0.005--0.013$\times$.

\paragraphX{Total write amplification}
Across the same five layers (\Cref{fig:ablation3}), the naive designs are counterproductive: the global-index and data-path layers push total WA to 1.62--1.69$\times$ on dedup-light workloads, which is \emph{above} NoDedup's 1.0$\times$, because metadata rewrite traffic exceeds what the thin duplicate stream saves. The full design inverts this: 0.15--0.34$\times$ on dedup-rich workloads and 0.81--0.85$\times$ even on dedup-light ones, below 1.0 across all five workloads (physical writes shrink 15.8--85.1\% while metadata adds only 1.4--2.2\%).

\subsection{Long-Run Behavior}
\label{sec:eval-long}

\begin{figure}[t]
\centering
\subcaptionbox{FSL (dedup-light).\label{fig:longrun-fsl}}{%
  \includegraphics[height=0.12\textheight]{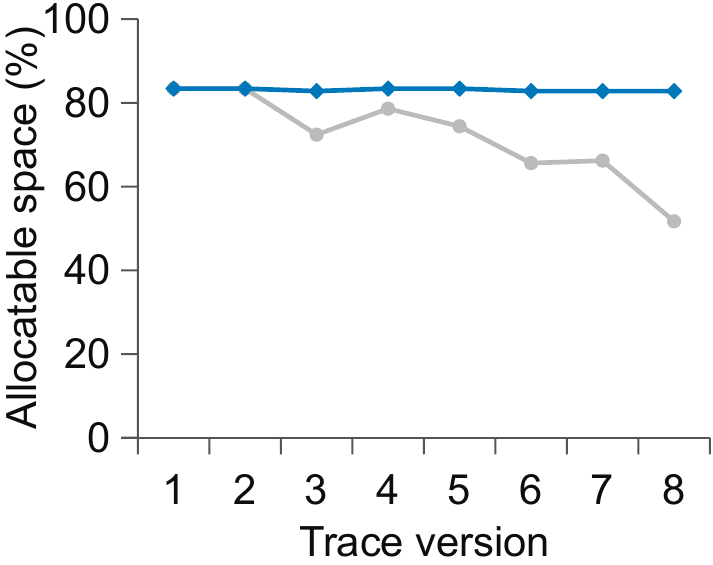}}
\subcaptionbox{Container (dedup-rich).\label{fig:longrun-container}}{%
  \includegraphics[height=0.12\textheight]{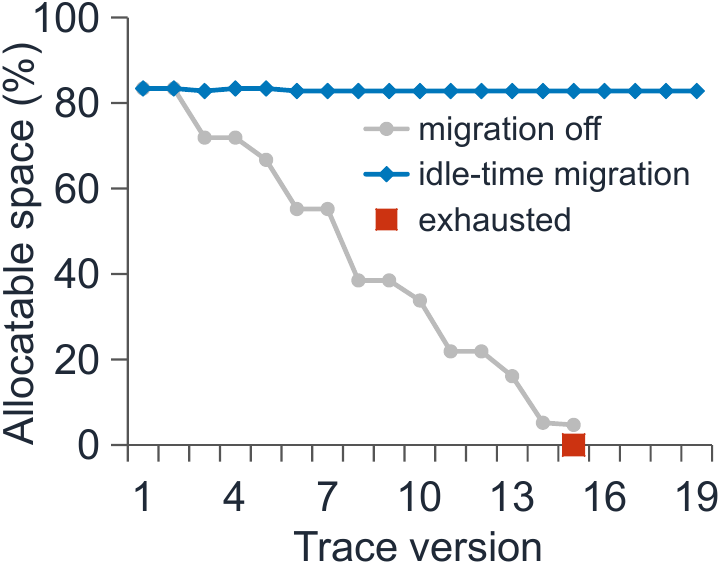}}%
\caption{Allocatable space over repeated 0.4\,M-write versions.}
\label{fig:longrun}
\end{figure}

\Cref{fig:longrun} validates cross-zone reference management over repeated 0.4\,M-write versions with aggressive expiry (192-zone device). With idle-time migration, allocatable space holds steady at $\sim$83\% across all 8 (FSL) and 19 (container) versions. Without migration, pinned zones accumulate monotonically. On container the device exhausts at 78\% of the stream. The incoming-count tracking and migration protocol thus prevent permanent capacity loss from cross-zone references.

\subsection{Sensitivity and Scalability}
\label{sec:eval-sensitivity}

\begin{figure}[t]
\centering
\subcaptionbox{Controller DRAM scalability.\label{fig:md-ra}}{%
  \includegraphics[height=0.128\textheight]{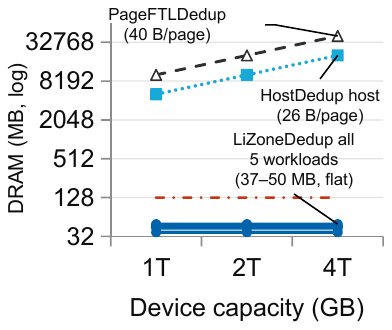}}%
\subcaptionbox{Zone size sensitivity.\label{fig:md-wamd}}{%
  \includegraphics[height=0.128\textheight]{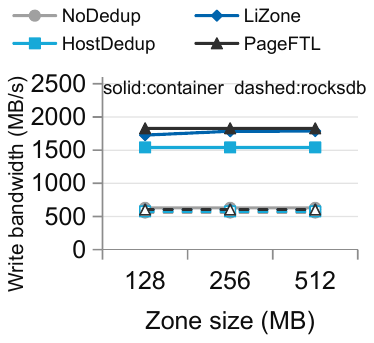}}%
\caption{Sensitivity and Scalability.}
\label{fig:sensitivity}
\end{figure}

\paragraphX{Controller DRAM scalability}
\sysname's controller DRAM is flat with device capacity on both workloads (\cref{fig:sensitivity}): it stays around 37--50\,MB from 1\,TB through 4\,TB. Capacity thus changes the zone count, never the index budget. The reason is zone-local indexing: each of the $\le$16 active zones maintains its own working index and filters in DRAM, while sealed zones' identity indexes reside on NAND and are fetched on demand. DRAM does grow as the caches fill---14.4\,MB after 1\,M writes on container, 29.8\,MB after 2\,M---but plateaus at the cache bound rather than growing with the data written. The uncached split-state design alone (+Split) runs in 2.7--8.4\,MB; on the full design the index proper (working index, filters, delta logs, and the route table) occupies 3.7--7.2\,MB and the reference plane 0.25\,MB, both flat with capacity. The DRAM cache is thus an optional accelerator that buys a 61$\times$ read-amplification reduction (1,396\,B to 22.7\,B per request on container) at a cost of 32.7--43.1\,MB.

\paragraphX{Zone size sensitivity}
Larger zones hold more unique pages, so duplicates are more likely to co-locate with their source (\Cref{fig:md-wamd}): on container, moving from 128\,MB to 256\,MB zones raises physical write reduction from 74.9\% to 77.4\% and write throughput by 3.4\%. On rocksdb nothing moves, because its duplicates are temporally adjacent. Zone size is thus a deployment knob rather than a correctness constraint: the default 128\,MB keeps per-zone indexes small ($\le$520\,KB at the observed maximum of 26,000 unique pages, fast to rebuild at seal); 256\,MB zones buy +2.5 points of reduction.

\begin{figure}[t]
\centering
\includegraphics[width=0.62\linewidth]{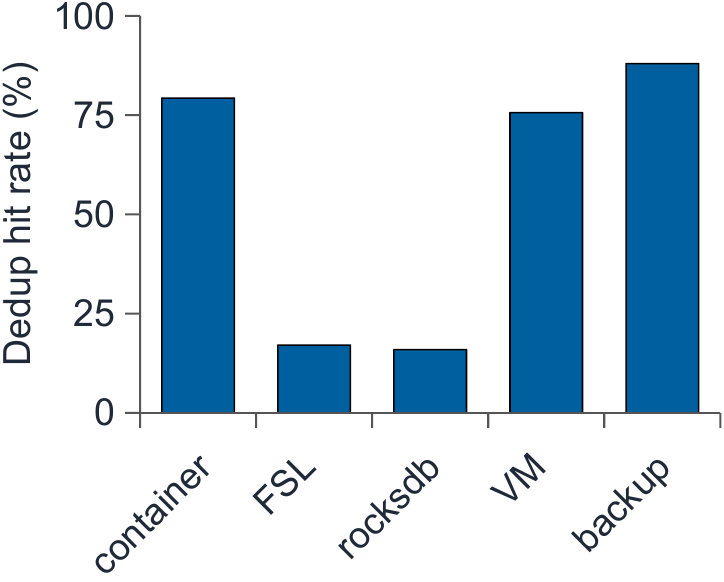}
\caption{FEMU cross-validation.}
\label{fig:femu-crossval}
\end{figure}

\paragraphX{FEMU Cross-Validation}
Absolute guest throughput on the FEMU prototype is capped by QEMU/TCG, so we validate data reduction rather than performance: as \Cref{fig:femu-crossval} shows, the dedup hit rate measured on FEMU agrees with the simulator on all five workloads. In write-heavy workloads where NAND program bandwidth is the bottleneck, suppressed writes free up channel/die parallelism, so the throughput improvement is structural, not a timing-model artifact. Together with the crash-injection results, this gives us confidence that the simulator faithfully models the dedup path, although no real ZNS controller hardware was available for end-to-end timing.


\section{Related Work}
\label{sec:related}

\paragraphX{ZNS systems and device-internal enhancements}
Performance characterizations~\cite{tehrany2022understanding,doekemeijer2023performance,tehrany2023contracts} establish the baseline contracts of contemporary ZNS devices. Prior work explores ZNS across the stack: RAID-like aggregation~\cite{kim2023raizn}, file-system deployment~\cite{seo2023f2fszns}, KV-store backends~\cite{oh2023zenfsplus,li2022lsmhybrid}, and block-layer indirection~\cite{im2022smallzone,choi2024zonegrouping}. DeZNS~\cite{jamil2024dezns} predicts duplicates with CRC32 and deduplicates offline in ZenFS, but stays above the block interface and cannot suppress writes at the device. Device-internal enhancements include wear optimization~\cite{long2024wazone}, GC offload~\cite{liu2024znscleaner,tan2024gczns}, variable zone sizes~\cite{wang2023flexzns}, fragmentation reduction~\cite{qi2022indef}, and orderless-write resolution~\cite{wang2024oz}; Tan et al.~\cite{tan2024gczns} use a small side-table for reference rewiring. None target deduplication. \sysname instead avoids device-side LBA-to-PPA re-binding by routing duplicates at index time.

\paragraphX{In-storage data reduction and the ZNS gap}
In-storage deduplication has a long history on page-mapped flash, where the FTL absorbs remapping cost~\cite{caftl2012,zou2022dedupelocality}. Designs span inline primary dedup~\cite{srinivasan2012idedup}, hybrid inline/out-of-line policies~\cite{li2014hybriddedup}, content-defined chunking with similarity routing~\cite{xia2016fastcdc,lillibridge2009sparse,xia2011silo}, and recent remap- or ECC-assisted schemes~\cite{zhou2021remapssd,wu2022ead} that still require either page-level remapping or a global fingerprint table. In-storage compression~\cite{wang2024balloonzns} is complementary. 

Host-side deduplication~\cite{khan2018clusterdedup,sun2018longterm} identifies duplicates before data reaches the device, trading host CPU and DRAM for write reduction. None fit ZNS's constraints: in-storage designs assume a page-level FTL or a multi-GB global hash table, neither of which fits ZNS's 64--256\,MB DRAM and sequential-write budgets; host-side designs cannot exploit last-moment write suppression or zero-host-cost fingerprinting.

\paragraphX{Foundational techniques}
Our design builds on three established techniques. First, approximate membership filters: counting Bloom filters~\cite{pandey2017countingfilter} support insert/delete for active zones, while static Xor and binary-fuse filters~\cite{graf2020xor,graf2022binaryfuse} achieve $\sim$8 bits/key for sealed zones---our lifecycle split exploits both. Second, append-friendly indexing: our shard merges borrow LSM-tree intuition~\cite{ren2017slimdb,sarkar2022lsmdesign} but operate at 4\,KB shard granularity, bounded by zone life cycle, with no multi-level cascade. Third, LSH routing: LSH has been used for similarity routing in deduplication~\cite{satuluri2011bayeslsh,shinde2010tcamlsh,shrivastava2017densification}, but \sysname is the first to deploy it as an in-storage zone router combined with an address-after-write primitive.


\section{Conclusion and Future Work}\label{sec:conclusion}
This paper presented \sysname, an in-storage deduplication system for ZNS SSDs that resolves the conflicts between ZNS's append-only semantics and conventional deduplication's mutable global mapping tables. By moving deduplication into the controller, it preserves ZNS invariants and enables cross-zone duplicate sharing via device-selected allocation, content-derived routing, and append-oriented metadata, cutting physical writes by up to 85\%, boosting write throughput by 1.1--3.5$\times$, and holding controller DRAM at 37--50\,MB regardless of capacity. Practical trade-offs remain: \texttt{dup\_append} requires NVMe vendor-specific extensions and host-stack modifications, and cross-zone references add host-side GC overhead. Future work includes hardware timing validation on real ZNS controllers, dynamic adaptation of $r$, pipelining cold-path descriptor fetches, and read-intensive workloads.

{\footnotesize
	\bibliographystyle{plainnat}%
	\bibliography{refs}%
}

\end{document}